\documentstyle[preprint,prl,aps]{revtex}

\begin{document}

\draft

\preprint{\vbox{\hfill SNUTP 97-033, UTHEP-358\\ \null\hfill
 March., 1997 }}

\narrowtext

\title{Comment on: ``Meson Masses in Nuclear Matter''}

\author{Tetsuo Hatsuda}
\address{Institute of Physics,
University of Tsukuba, Tsukuba, Ibaraki 305, Japan}
\author{Su Houng Lee}
\address{Department of Physics, Yonsei University, Seoul, 120-749, Korea}
%
\maketitle

\bigskip

\bigskip


\bigskip
\bigskip

 In a recent letter \cite{EI}, Eletsky and Ioffe 
 estimated the energy shift of 
 $\rho$-meson  with {\em high} momentum (${\bf q}$) in nuclei 
 using the Glauber approximation. They
  also made remarks on the {\em low} ${\bf q}$
 vector mesons in medium. In this Comment,
 we will discuss the second point and show that
 (i)  mean-field  physics for the mass shift of
 low-${\bf q}$ mesons is  overlooked in \cite{EI},
 and (ii) the {\em short distant} operator product expansion
  (OPE) is relevant to analyze the low-${\bf q}$ mesons
 in QCD sum rules  contrary to the claim in \cite{EI}.

  Mesons in  nuclei have 
 considerable multiple scattering
 when  $|{\bf q}|^{-1} \gg d$ ($d$ is the inter-nucleon distance).
 The relevant quantity here is  known to be
 a  polarization function (generalized optical potential)
 $\Pi(\omega, {\bf q})$  \cite{EW},
  which is written, up to the Lorentz-Lorenz correction, as
\begin{eqnarray}
\label{eq1}
\Pi(\omega, {\bf q}) = - 16 \pi 
 \int {d^3 {\bf p} \over (2 \pi)^3} F(\omega,{\bf q};{\bf p}) n_{_F}({\bf p}),
\end{eqnarray}  
where $n_{_F}$ is the fermi distribution,
 $F$ is an {\em in-medium} forward scattering amplitude of an incoming  
 vector-meson (V)
  with a nucleon (N) of momentum ${\bf p}$.
   At extremely low density,  eq.(\ref{eq1}) reduces to 
 $\Pi(\omega, {\bf q}) = - 4 \pi F(\omega,{\bf q};0) \rho$,
  which  does not however have practical use, since 
 $F$ has not only a smooth ${\bf p}$-dependence
 (from e.g. the $t$-channel meson exchanges considered in \cite{FS})
 but also a possible rapid ${\bf p}$-dependence from  
  s-channel baryon-resonances coupled to the $V$-$N$ system.
 In other words, the averaging of $F$ over
 the fermi sea is essential 
 for the mass shift at low ${\bf q}$.
  Note also that
 low-${\bf q}$ mesons probe
 large distances and hence the averaged nuclear properties. 
 The mean-field (MF) + RPA treatment of $\Pi$, which has been
 employed in hadronic model calculations
  (see e.g. \cite{JPW}), is most suited for studying 
 these features.
  The importance of this MF description
 for low-${\bf q}$ mesons  is overlooked in  \cite{EI}. 

  The MF description is also 
  compatible with the finite-density QCD sum rules \cite{HL}.
 Consider the vector-current correlation in medium,
\begin{eqnarray}
D_{\mu\nu} (Q^2, Q^2/q\cdot P) = i \int d^4x e^{iqx}\langle P |
      T [ J_\mu(x) J_\nu(0) ] | P \rangle \ ,
\end{eqnarray}
 with 
 $Q^2 \equiv- q^2$ and   
$ |P \rangle $ being the nuclear ground state
 with momentum $P^{\mu}$. At large $Q^2$, 
 one can make OPE in two ways \cite{Wilson}:
 the short-distance (SD) 
expansion where $(Q^2, Q^2/(q \cdot P)^2) \rightarrow 
 (\infty,{\rm finite})$, 
 and the light-cone (LC)  expansion where $(Q^2, Q^2/q \cdot P ) 
 \rightarrow  (\infty, {\rm finite})$.   
 Both are at best an asymptotic expansion in $1/Q^2$ with 
different composite operators  contributing at each order:  The 
 SD (LC) expansion is dictated by the
 dimension (twist) of the operators.
 Since we are interested in ${\bf q} \simeq 0$ with 
 ${\bf P}= 0$,  OPE in the deep-euclidian region 
 ($-\omega^2 \rightarrow \infty$) is obtained from 
 the SD expansion by re-expanding $1/Q^2$  and  
 $(q \cdot P)^2/Q^2$ 
 in terms of  $1/\omega^2$,    
 which  is a natural generalization 
 of the QCD sum rules in the vacuum \cite{HL,Griegel}.
  More importantly, from the analytic properties of
 $D_{\mu \nu}$, one can relate the OPE 
  to the spectral density via 
 the well-known fixed ${\bf q}$ dispersion relation \cite{FW}  
\begin{eqnarray}
\label{disp1}
{\rm Re}\  D(\omega^2 < 0, {\bf q})
=\int_0^\infty dx^2 {\rho(x,{\bf q}) \over (x^2-\omega^2)}
 + ({\rm subtraction}).
\end{eqnarray}
 As in the vacuum sum rules,
 we  keep only the contributions up 
 to dimension 6 operators.  The actual expansion parameter
 at low density is $\Lambda_{_{QCD}}^l p_F^m |{\bf q}|^n/\omega^{l+n+m}$
 ($p_F$ is the fermi momentum).
 Density dependent contributions from  
 dimension 4 and dimension 6 operators have been
  shown not to spoil the asymptotic nature of the series \cite{HL}.
   In the work by Drukarev and Levin \cite{DL92},
 a kinematics  $Q^2 \rightarrow \infty$ with
 $s = (q+P/A)^2$ 
 fixed ($A$ is the number of nucleons) or its variant are adopted.
 This leads to the LC expansion as quoted in 
 \cite{EI}. However,  a dispersion relation for $Q^2$ with fixed $s$
 is {\em assumed} in \cite{DL92}. 
 Such a dispersion relation,  unlike eq.(\ref{disp1}),  
 cannot be derived from analytic properties
 of $D_{\mu \nu}$ alone \cite{Griegel}.  Also fixing $s$ 
 inevitably gives a large ${\bf q}$ limit, which is inappropriate
 to extract the MF physics we are interested in.


\begin{references}

\bibitem{EI} V. L. Eletsky and B. L. Ioffe, 
 Phys. Rev. Lett. {\bf 78}, 1010 (1997).
 
\bibitem{EW} 
  A. B. Migdal, in {\em Mesons in Nuclei}, ed. M. Rho and D. Wilkinson 
 (North-Holland, Amsterdam, 1979).
  T. Ericson and W. Weise, {\em Pions and Nuclei},
  (Oxford Univ. Press, Oxford, 1988).

\bibitem{FS} B. Friman and M. Soyeur, 
 Nucl. Phys. {\bf A600}, 477 (1996).


\bibitem{JPW} H. C. Jean, J. Piekarewicz and A. G. Williams, 
 Phys. Rev. {\bf C49}, 1981 (1994). 

\bibitem{HL} T. Hatsuda and Su H. Lee, Phys. Rev. {\bf C46}, R34 (1992).
 T. Hatsuda, Su H. Lee and H. Shiomi, Phys. Rev. {\bf C52}, 3364 (1995).

\bibitem{Wilson} 
T. Muta, {\em Foundations of Quantum Chromodynamics}, (World-Scientific,
 Singapore, 1987).

\bibitem{FW} A. L. Fetter and J. D. Walecka, {\em Quantum Theory of
 Many-Particle Systems}, (McGraw-Hill, New York, 1971).

\bibitem{Griegel}  T. D. Cohen, R.J. Furnstahl, D.K. Griegel and X.M. Jin,
  Prog. in Part. Nucl. Phys. {\bf 35}, 221 (1995), section 3.7.
 D. K. Griegel, PhD Thesis (Univ. of Maryland) 1991. 



\bibitem{DL92} E.G. Drukarev and E.M. Levin, Prog. Part. Nucl. Phys. {\bf 27}, 
 77 (1991).

\end{references}
\end{document}